\documentclass[sigconf]{acmart}

\copyrightyear{2026}
\acmYear{2026}
\setcopyright{cc}
\setcctype{by}
\acmConference[ICS '26]{2026 International Conference on Supercomputing}{July 06--09, 2026}{Belfast, United Kingdom}
\acmBooktitle{2026 International Conference on Supercomputing (ICS '26), July 06--09, 2026, Belfast, United Kingdom}
\acmDOI{10.1145/3797905.3807833}
\acmISBN{979-8-4007-2522-7/2026/07}

\usepackage{subfigure}
\usepackage{caption}

\usepackage{enumitem} 
\usepackage{amsthm}

\usepackage{multirow}
\usepackage{makecell}

\usepackage[ruled,vlined, linesnumbered]{algorithm2e}

\usepackage{comment}

\usepackage{pifont} 

\newcommand{\circled}[1]{%
\raisebox{-0.15ex}{\scalebox{1.2}{\ding{\the\numexpr 181+#1\relax}}}}
\newcommand{\circledwhite}[1]{%
\raisebox{-0.15ex}{\scalebox{1.2}{\ding{\the\numexpr 171+#1\relax}}}}

\usepackage{cleveref}
\Crefname{figure}{Fig.}{Figs.}
\Crefname{section}{Sec.}{Secs.}
\Crefname{table}{Tab.}{Tabs.}
\Crefname{algorithm}{Alg.}{Algs.}

\newcommand{\nbc}[3]{
  {\colorbox{#3}{\bfseries\sffamily\scriptsize\textcolor{white}{#1}}}
  {\textcolor{#3}{\sf\small$\blacktriangleright$\textit{#2}$\blacktriangleleft$}}
}

\definecolor{babcolor}{rgb}{0.9,0.45,0.1}
\newcommand{\bab}[1]{\nbc{BAB}{#1}{babcolor}}

\newcommand{\method}{\texttt{SHIRO}}
\usepackage{microtype}

\usepackage{booktabs}   

\begin{document}

\title{SHIRO: Near-Optimal Communication Strategies for Distributed Sparse Matrix Multiplication}

\author{Chen Zhuang}
\orcid{0009-0006-4156-9879}
\affiliation{\institution{Institute of Science Tokyo}
\city{Tokyo}
\country{Japan}}
\affiliation{\institution{RIKEN Center for Computational Science}
\city{Kobe}
\country{Japan}}
\email{chen.zhuang@riken.jp}

\author{Lingqi Zhang}
\orcid{0000-0002-2452-1551}
\affiliation{\institution{RIKEN Center for Computational Science}
\city{Kobe}
\country{Japan}}
\email{lingqi.zhang@riken.jp}

\author{Benjamin Brock}
\orcid{0000-0003-1488-1622}
\affiliation{\institution{Intel Corporation}
\city{San Francisco}
\country{USA}}
\email{benjamin.brock@intel.com}

\author{Du Wu}
\orcid{0000-0002-4002-0837}
\affiliation{\institution{Institute of Science Tokyo}
\city{Tokyo}
\country{Japan}}
\affiliation{\institution{RIKEN Center for Computational Science}
\city{Kobe}
\country{Japan}}
\email{du.wu@riken.jp}

\author{Peng Chen}
\orcid{0000-0003-1244-3151}
\affiliation{\institution{RIKEN Center for Computational Science}
\city{Kobe}
\country{Japan}}
\email{peng.chen@riken.jp}

\author{Toshio Endo}
\orcid{0000-0001-7297-6211}
\affiliation{\institution{Institute of Science Tokyo}
\city{Tokyo}
\country{Japan}}
\email{endo@scrc.iir.isct.ac.jp}

\author{Satoshi Matsuoka}
\orcid{0000-0003-1910-8532}
\affiliation{\institution{RIKEN Center for Computational Science}
\city{Kobe}
\country{Japan}}
\affiliation{\institution{Institute of Science Tokyo}
\city{Tokyo}
\country{Japan}}
\email{matsu@acm.org}

\author{Mohamed Wahib}
\orcid{0000-0002-7165-2095}
\affiliation{\institution{RIKEN Center for Computational Science}
\city{Kobe}
\country{Japan}}
\email{mohamed.attia@riken.jp}

\renewcommand{\shortauthors}{Zhuang et al.}

\begin{abstract}

Distributed Sparse Matrix-Matrix Multiplication (SpMM) is a fundamental operation in high-performance computing and deep learning applications. The major performance bottleneck in distributed SpMM lies in substantial communication overhead, which limits both performance and scalability. In this paper, we identify two key sources of communication inefficiency in distributed SpMM: redundant data transfer due to sparsity unawareness, and suboptimal utilization of hierarchical network topology. To address these, we propose (1) a fine-grained, sparsity-aware communication strategy that reduces communication overhead by exploiting the sparsity pattern of the sparse matrix, and (2) a hierarchical communication strategy that maps the sparsity-aware strategy onto two-tier GPU network architectures, minimizing redundant data movement across slower inter-node links. We implement these optimizations in a comprehensive distributed SpMM framework, \method{}. Extensive evaluations on real-world datasets show that \method{} demonstrates strong scalability up to 128 GPUs, achieving geometric mean speedups of 221.5$\times$, 56.0$\times$, 23.4$\times$, and 8.8$\times$ in SpMM over four state-of-the-art baselines (CAGNET, SPA, BCL, and CoLa, respectively) at this scale.


\end{abstract}



\maketitle




\vspace{-10pt}
\section{Introduction}

Sparse matrix-dense matrix multiplication (SpMM) is a fundamental operation across numerous computational domains. In high-performance computing, SpMM serves as a critical building block for applications including graph algorithms~\cite{tiskin2001all, brock2024rdma} and block iterative solvers~\cite{o1980block, grimes1994shifted, sadkane1993block, gutknecht2007block}. In deep learning, SpMM has become the core computational kernel in Graph Neural Networks (GNNs), where it implements message-passing operations~\cite{gilmer2017neural} in widely used frameworks such as PyTorch Geometric~\cite{fey2019fast} and Deep Graph Library~\cite{wang2019deep}. Beyond these domains, SpMM also plays an important role in recommendation systems~\cite{he2017neural} and knowledge graphs~\cite{wang2017knowledge}.


Beyond its broad applicability, which motivates extensive single-node SpMM optimizations~\cite{gale2020sparse, ye2023sparsetir, fan2024dtc, wang2023tc, pang2024row}, optimizing distributed SpMM is essential for handling sparse matrices that exceed single-processor memory capacity. 
For example, finite element simulations generate matrices with millions of degrees of freedom~\cite{schenk2004solving}, and graph analytics on social networks and web graphs process adjacency matrices containing billions of nodes and edges~\cite{hu2020ogb, snapnets, leskovec2016snap}.
Such scale necessitates efficient distributed SpMM implementations.


Achieving this efficiency, however, is hindered by communication overhead.
As input matrices are distributed across processors, each processor must fetch remote matrix blocks during computation~\cite{koanantakool2016communication, block2024two, acer2016improving, abubaker2024spcomm3d}. This bottleneck is particularly severe in strong-scaling scenarios such as distributed GNN training, where communication consumes up to 83.3\% of execution time on 100 NVIDIA V100 GPUs~\cite{tripathy2020reducing} and 85.7\% on 192 GPUs~\cite{brock2024rdma}. This problem intensifies in modern GPU clusters where computational throughput far exceeds inter-node communication bandwidth~\cite{brock2024rdma, tripathy2020reducing, mukhopadhyay2024sparsity, selvitopi2021distributed}.

Extensive research has been conducted to reduce communication overhead in distributed SpMM~\cite{koanantakool2016communication, mukhopadhyay2024sparsity, zhang2025cola, tripathy2020reducing, bharadwaj2022distributed, selvitopi2021distributed, acer2016improving, brock2024rdma}. {However}, existing methods still suffer from communication inefficiencies at two levels:

\noindent\textbf{(1) Redundancy caused by communication strategies.} Existing communication approaches can be broadly categorized into two types: (i) sparsity-oblivious methods~\cite{tripathy2020reducing, bharadwaj2022distributed, selvitopi2021distributed, brock2024rdma} transfer entire dense matrix blocks regardless of the actual sparsity patterns, transmitting unnecessary data that introduces significant communication overhead;
(ii) sparsity-aware methods~\cite{mukhopadhyay2024sparsity, zhang2025cola, acer2016improving, abubaker2024spcomm3d} attempt to reduce this overhead by transferring data based on the actual demand determined by the sparse matrix patterns. However, these sparsity-aware approaches still suffer from communication redundancy as they only consider partial sparsity information, failing to fully exploit the complete sparsity pattern.

\noindent {\textbf{(2) Flat communication over a hierarchical network.} Most existing distributed SpMM methods employ flat communication patterns that assume uniform costs across all process pairs. However, modern computing clusters, especially GPU-accelerated ones, exhibit hierarchical networks where bandwidth varies significantly across levels. Such hierarchy-unaware communication causes redundant data transfers, as processes within the same fast-link group independently fetch identical data through slow inter-group connections. Although recent work~\cite{zhang2025cola, bienz2018reducing, bienz2020reducing, lockhart2023performance} partially addresses this inefficiency, it supports only limited sparsity-aware communication strategies that partially reduce redundancy.}

To address these inefficiencies, 
we propose \method{}\footnote{\url{https://github.com/zhuangbility111/shiro-spmm}}, a communication-efficient distributed framework that jointly optimizes sparsity-aware communication strategies and exploits hierarchical network topology to improve the scalability and performance of distributed SpMM. 
Our main contributions are:
\begin{itemize}[leftmargin=1em]
    \item \textbf{Joint row-column sparsity-aware communication strategy.} We propose a joint sparsity-aware communication strategy that comprehensively exploits complete sparsity patterns to minimize communication redundancy. We analyze the relationship between sparsity patterns and communication requirements, formulate 
    the optimal communication volume problem as a minimum weighted covering problem {on sparse matrices}, and provide an optimal polynomial-time algorithm based on graph theory.
    \item 
    \textbf{{Row-column sparsity-aware} hierarchical communication strategy.
    } 
    {We present a hierarchical communication strategy that integrates the proposed sparsity-aware strategy with hierarchical network topologies. We first design a group-based communication scheme to minimize communication over bandwidth-limited inter-group links, and further introduce an overlapping scheduling strategy that exploits the complementary nature of intra- and inter-group transfers in our sparsity-aware strategy to fully utilize bandwidth at both levels.}
    \item Comprehensive experimental results on real-world datasets demonstrate that \method{} achieves up to 96.3\% communication volume reduction and exhibits strong scalability on up to 128 GPUs, delivering geometric mean speedups of 221.5$\times$, 56.0$\times$, 23.4$\times$, and 8.8$\times$ {in SpMM} over four state-of-the-art baselines at this scale.
    {In addition, by applying \method{} to large-scale GNN training, our framework achieves up to 1.78$\times$ end-to-end speedup over PyTorch Geometric with less than 13.2\% preprocessing overhead.}
\end{itemize}

\begin{figure*}[ht]
    \centering
    \subfigure[1D partition distributed SpMM. Dark regions: off-diagonal submatrices for SpMM on remote data.
    ]{
        \begin{minipage}[b]{0.38\textwidth}
            \includegraphics[width=1\textwidth]{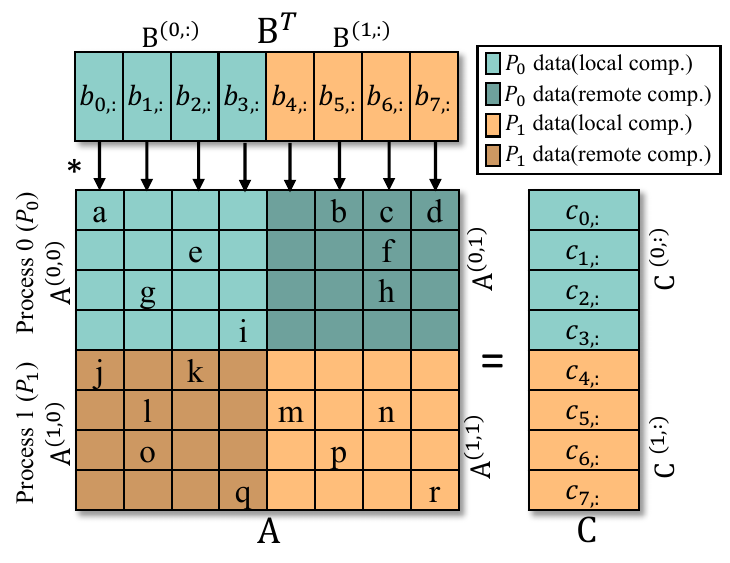}
            \vspace{-15pt}
        \end{minipage}
        \label{fig/background/distributed_spmm}
    }
    \hspace{3mm}
    \subfigure[Sparsity-aware (column-based) distributed SpMM from Process 0's view. $P_0$ fetches required $\mathbf{B}$ rows from $P_1$ based on unique column indices of nonzeros in $\mathbf{A}^{(0,1)}$. Communication volume is 3 rows.
    ]{
        \begin{minipage}[b]{0.4\textwidth}
        \includegraphics[width=1\textwidth]{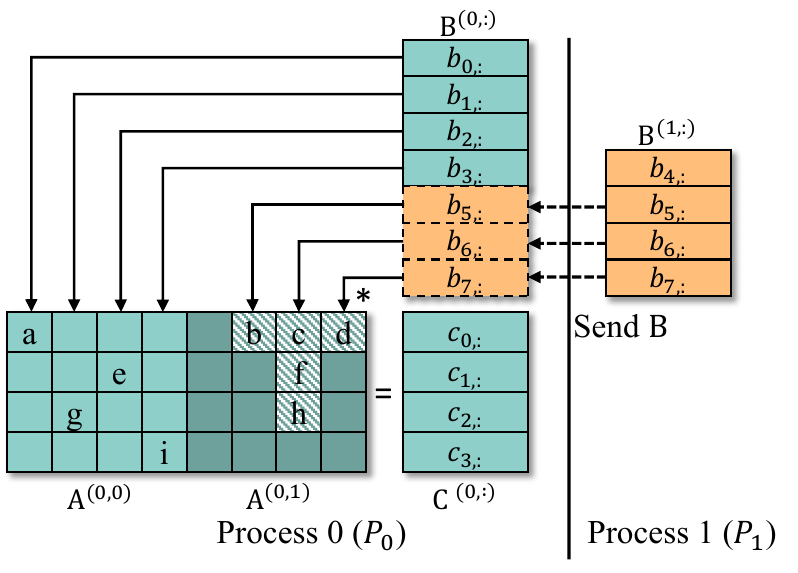}
        \vspace{-15pt}
        \end{minipage}
    \label{fig/method:complicated_post_graph}
    } 
    \\
    \vspace{-10pt}
    \subfigure[Sparsity-aware (row-based) distributed SpMM from Process 0's view. $P_1$ computes partial results and sends required $\mathbf{C}$ rows to $P_0$ based on unique row indices in $\mathbf{A}^{(0,1)}$. Communication volume is 3 rows.
    ]{
        \begin{minipage}[b]{0.38\textwidth}
            \includegraphics[width=1\textwidth]{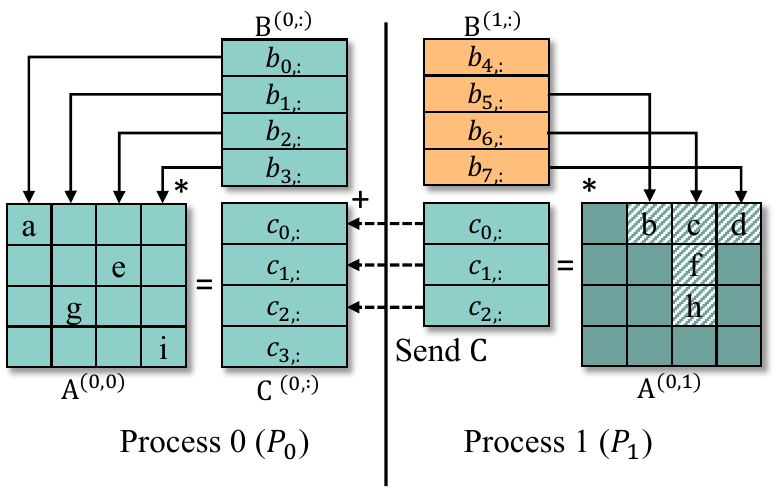}
            \vspace{-15pt}
        \end{minipage}
        \label{fig/method:complicated_pre_graph}
    } 
    \hspace{2.5mm}
    \subfigure[Our proposed joint row-column distributed SpMM from Process 0's view. $P_1$ computes partial results and sends required $\mathbf{B}$ and $\mathbf{C}$ rows to $P_0$. Communication volume reduced from 3 to 2 rows.
    ]{
        \begin{minipage}[b]{0.45\textwidth}
            \includegraphics[width=1\textwidth]
            {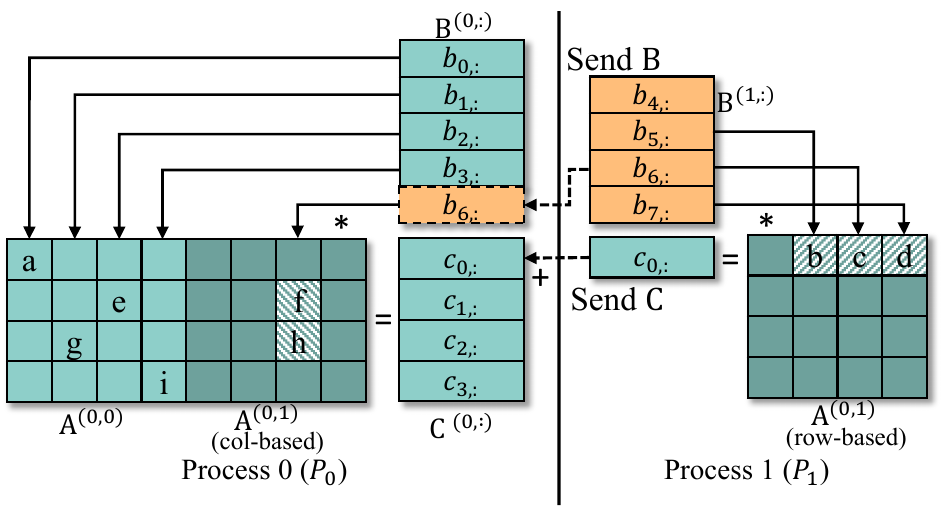}
            \vspace{-15pt}
        \end{minipage}
        \label{fig/method:complicated_pre_post_graph}
    }  
    \vspace{-13pt} 
    \caption{Different communication strategies for 1D distributed SpMM {(illustrating communication from process $P_1$ to $P_0$).}}
    \vspace{-10pt}
    \label{fig/method:complicated_graph}
\end{figure*}

\begin{table}[t]
\centering
\caption{{Notation used in the distributed SpMM communication analysis.}}
\label{tab:notation_comm}
\vspace{-5pt}
\footnotesize
\begin{tabular}{@{}l p{0.74\linewidth}@{}} 
\toprule
\textbf{Notation} & \textbf{Description} \\
\midrule
$\mathbf{A}^{(p,:)}$, $\mathbf{B}^{(p,:)}$, $\mathbf{C}^{(p,:)}$ & Row blocks of matrices $\mathbf{A}, \mathbf{B}$, and $\mathbf{C}$ owned by process $p$ \\
$\mathbf{A}^{(p,q)}$ & Sub-block of $\mathbf{A}$ at block row $p$ and block column $q$ \\
$a_{ij}$ & The $(i,j)$-th element of matrix $\mathbf{A}$ \\
$\mathbf{a}_{i,:}, \mathbf{a}_{:,j}$ & The $i$-th row and $j$-th column of matrix $\mathbf{A}$ \\
$M$ & Number of rows in $\mathbf{A}^{(p,:)}$ and $\mathbf{C}^{(p,:)}$ \\
$K$ & Number of columns in $\mathbf{A}^{(:,p)}$ and rows in $\mathbf{B}^{(p,:)}$ \\
$N$ & Number of columns in $\mathbf{B}^{(p,:)}$ and $\mathbf{C}^{(p,:)}$ \\
$\mathit{sz}_{\mathrm{dt}}$ & Size of each data element (in bytes) \\
$\mu$ & Size of the minimum vertex cover set \\
{$\mathrm{Cols}(\mathbf{A}^{(p,q)})$} & Unique column indices of nonzeros in $\mathbf{A}^{(p,q)}$ \\
{$\mathrm{Rows}(\mathbf{A}^{(p,q)})$} & Unique row indices of nonzeros in $\mathbf{A}^{(p,q)}$ \\
\bottomrule
\vspace{-20pt}
\end{tabular}
\end{table}

\section{Background}
\subsection{Distributed SpMM}
Sparse Matrix-Matrix Multiplication (SpMM) is a fundamental operation where a sparse matrix $\mathbf{A}$ is multiplied by a dense matrix $\mathbf{B}$ to produce a dense result $\mathbf{C}$: $\mathbf{C} = \mathbf{A}\mathbf{B}$. 
For large-scale matrices, distributed SpMM addresses memory constraints and improves performance, where matrices $\mathbf{A}$, $\mathbf{B}$, and $\mathbf{C}$ are partitioned into multiple blocks and distributed across different processors. Each processor must communicate with others to retrieve the necessary remote matrix blocks for computing its assigned portion of the result matrix. This paper focuses on 1D row-partitioned distributed SpMM, where all matrices are divided into row blocks and distributed across processors, which is widely adopted in practical applications such as Graph Neural Networks (GNNs) due to its simplicity and effectiveness, as well as its compatibility with sparsity-aware methods, which can exploit the skewed nonzero distributions observed in many datasets.
{While we focus on the 1D row partitioning strategy, our proposed method applies to any distributed SpMM setup that requires exchanging blocks of the dense matrix $\mathbf{B}$ or $\mathbf{C}$ among processors, regardless of the underlying partitioning strategy.}




\subsection{1D Row-partitioned Distributed SpMM}
In 1D row-partitioned distributed SpMM, matrices $\mathbf{A}$, $\mathbf{B}$, and $\mathbf{C}$ are partitioned along rows into different row blocks and distributed across processes. As illustrated in~\Cref{fig/background/distributed_spmm}, matrices $\mathbf{A}$, $\mathbf{B}$, and $\mathbf{C}$ are partitioned into $\mathbf{A}^{(0,:)}$ and $\mathbf{A}^{(1,:)}$, $\mathbf{B}^{(0,:)}$ and $\mathbf{B}^{(1,:)}$, and $\mathbf{C}^{(0,:)}$ and $\mathbf{C}^{(1,:)}$, which are allocated to processes $P_0$ and $P_1$, respectively. Under this partitioning scheme, process $P_0$ is responsible for computing $\mathbf{C}^{(0,:)}$ using $\mathbf{A}^{(0,:)}$ and the complete matrix $\mathbf{B}$, while process $P_1$ computes $\mathbf{C}^{(1,:)}$ using $\mathbf{A}^{(1,:)}$ and the complete matrix $\mathbf{B}$.

In 1D row-partitioned SpMM, each process's sparse matrix block can be divided based on data locality. Taking process $P_0$ as an example (\Cref{fig/background/distributed_spmm}), its sparse matrix $\mathbf{A}^{(0,:)}$ consists of the diagonal block $\mathbf{A}^{(0,0)}$ that requires only local dense matrix $\mathbf{B}^{(0,:)}$, and the off-diagonal block $\mathbf{A}^{(0,1)}$ that requires remote dense matrix $\mathbf{B}^{(1,:)}$ from process $P_1$. This locality pattern naturally leads to a four-stage execution: (1) \textbf{Local computation}, where $P_0$ computes with $\mathbf{A}^{(0,0)}$ and local $\mathbf{B}^{(0,:)}$; (2) \textbf{Communication}, where $P_0$ retrieves remote $\mathbf{B}^{(1,:)}$ from $P_1$; (3) \textbf{Remote computation}, where $P_0$ computes with $\mathbf{A}^{(0,1)}$ and the received $\mathbf{B}^{(1,:)}$; and (4) \textbf{Result aggregation}, where partial results are summed to obtain final $\mathbf{C}^{(0,:)}$. As the system scales to more processors, the \textit{local computation} decreases while \textit{communication} increases {(as each processor's local block $\mathbf{B}^{(0,:)}$ shrinks proportionally while the aggregate size of remote blocks $\mathbf{B}^{(1,:)},\mathbf{B}^{(2,:)}\ldots \mathbf{B}^{(n,:)}$ grows proportionally with {the number of }processors)}, shifting the primary bottleneck to communication.

\section{Motivation} \label{sec:motivation}
This section presents our motivation for reducing communication overhead {in distributed SpMM}. We examine why existing approaches fail to minimize this overhead despite various \mbox{optimization} attempts. Our analysis reveals two fundamental issues: current communication strategies transmit redundant data {(\Cref{sec:mot_strategies})}, and they ignore the hierarchical structure of modern interconnects {(\Cref{sec:mot_hierarchy})}. {We demonstrate that these problems are inter-dependent: solving one without the other yields limited benefits, motivating our unified approach (\Cref{sec:mot_summary}).}

\subsection{Redundancy in Existing Communication Strategies}\label{sec:mot_strategies}
In 1D row-partitioned distributed SpMM, each process $p$ owns row blocks $\mathbf{A}^{(p,:)}$, $\mathbf{B}^{(p,:)}$, and $\mathbf{C}^{(p,:)}$. To compute its output $\mathbf{C}^{(p,:)}$, each process obtains non-local data by either receiving required rows of $\mathbf{B}$ or collecting partial results of $\mathbf{C}$ from other processes. 
We examine three representative communication schemes for distributed SpMM, identifying redundancy reduction opportunities. \Cref{tab:notation_comm} summarizes the notations used in our analysis.

\noindent\textit{\textbf{1) Sparsity-oblivious (block-based) communication:}}
{In the sparsity-oblivious scheme~\cite{tripathy2020reducing}, each process fetches the complete row blocks $\mathbf{B}^{(q,:)}$ from each remote process $q$, regardless of the sparsity pattern in $\mathbf{A}$.}
The communication volume from process $q$ to process $p$ equals the size of the row block $\mathbf{B}^{(q,:)}$:
\begin{equation}\footnotesize
    V^{q,p}_\mathrm{block} = K \cdot N \cdot \mathit{sz}_{\mathrm{dt}}
\end{equation}

\noindent\textit{\textbf{2) Sparsity-aware (column-based) communication:}}
The sparsity-aware scheme~\cite{mukhopadhyay2024sparsity, zhang2025cola} fetches only the rows of $\mathbf{B}$ corresponding to unique column indices of nonzeros in $\mathbf{A}$.
As shown in~\Cref{fig/method:complicated_post_graph}, process $P_0$ fetches only rows $\mathbf{b}_{5,:}$, $\mathbf{b}_{6,:}$, and $\mathbf{b}_{7,:}$ from $P_1$ based on the nonzero columns in $\mathbf{A}^{(0,1)}$.
The communication volume from process $q$ to process $p$ is:
\begin{equation}\footnotesize
    V^{q,p}_{\mathrm{col}} = \bigl|\mathrm{Cols}(\mathbf{A}^{(p,q)})\bigr| \cdot N \cdot \mathit{sz}_{\mathrm{dt}}
\end{equation}

\noindent\textit{\textbf{3) Sparsity-aware (row-based) communication:}}
The row-based strategy transfers partial results of $\mathbf{C}$ based on row indices of nonzeros in $\mathbf{A}$, instead of transferring $\mathbf{B}$ rows based on column indices.
As shown in~\Cref{fig/method:complicated_pre_graph}, $P_1$ computes partial results for rows $\mathbf{c}_{0,:}$, $\mathbf{c}_{1,:}$, and $\mathbf{c}_{2,:}$ corresponding to nonzero rows in $\mathbf{A}^{(0,1)}$ and sends them to $P_0$.
The communication volume from process $q$ to process $p$ is:
\begin{equation}\footnotesize
    V^{q,p}_{\mathrm{row}} = \bigl|\mathrm{Rows}(\mathbf{A}^{(p,q)})\bigr| \cdot N \cdot \mathit{sz}_{\mathrm{dt}}
\end{equation}

\noindent\textit{\textbf{4) Analysis of Communication Redundancy:}}
Existing schemes fail to fully exploit the sparsity structure, leaving significant communication redundancy. The block-based approach ignores sparsity entirely, transferring $K \cdot N$ data regardless of {the specific data required by $\mathbf{A}$'s sparsity pattern}. The sparsity-aware schemes improve upon this by exploiting sparsity in a single dimension---reducing communication to {the number of unique columns $|\mathrm{Cols}(\mathbf{A}^{(p,q)})| \cdot N$ (column-based) \emph{or} the number of unique rows $|\mathrm{Rows}(\mathbf{A}^{(p,q)})| \cdot N$ (row-based)}. However, {as we will demonstrate}, single-dimension optimization still leaves substantial redundancy. As illustrated in~\Cref{fig/method:complicated_pre_post_graph}, a joint approach that considers both dimensions can dramatically improve efficiency---for the sparse block $\mathbf{A}^{(0,1)}$, only 2 rows ($\mathbf{b}_{6,:}$ and $\mathbf{c}_{0,:}$) are needed compared to 3 rows for single-dimension approaches. This gap widens for sparser matrices with more complex patterns, motivating a joint optimization framework that simultaneously exploits sparsity in both dimensions.



\subsection{Flat Communication over Hierarchical Networks}\label{sec:mot_hierarchy}

\begin{figure}[t!]
    \centering
    \includegraphics[clip,width=0.5\textwidth]{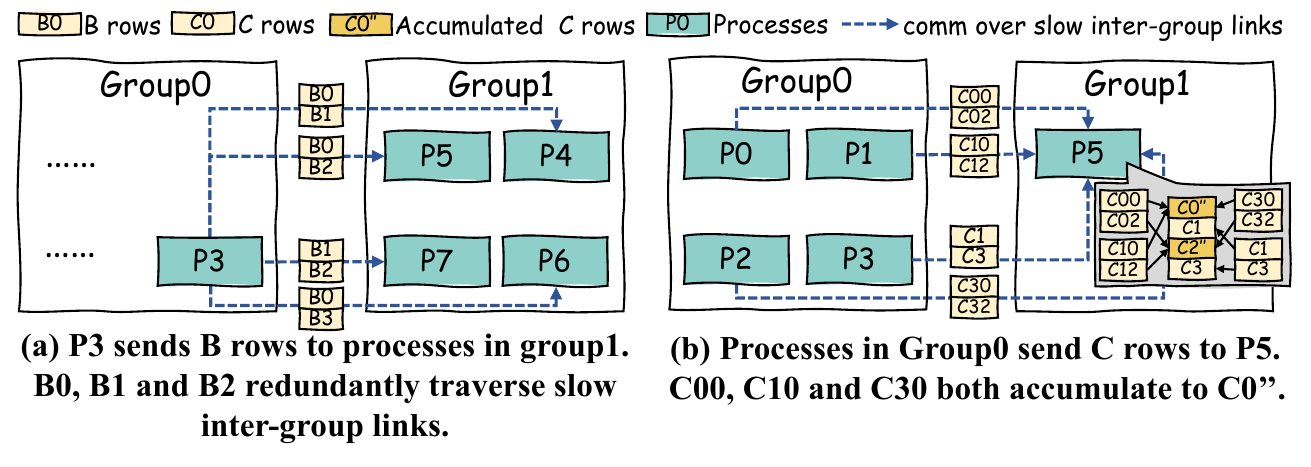}
    \vspace{-20pt}
    \caption{Redundant transfers over slow inter-group links.}
    \vspace{-15pt}
    \label{fig/methods/motivation_hierarchical_comm}
\end{figure}

\begin{figure*}[t!]
    \centering
    \includegraphics[width=1.02\textwidth]{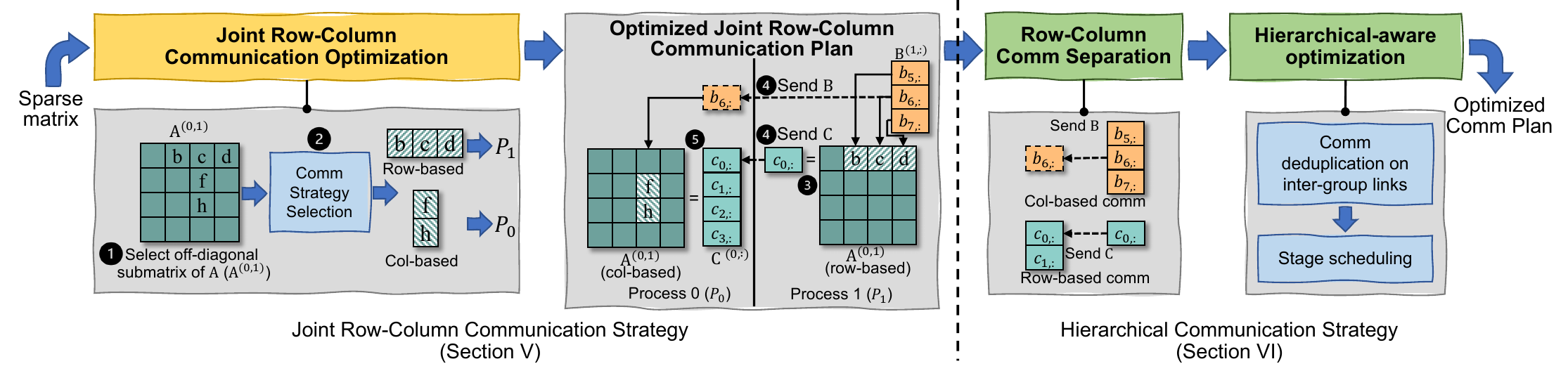}
    \vspace{-24pt}
    \caption{Overview of our proposed optimization {strategies}.}
    \vspace{-10pt}
    \label{fig/methods/overview}
\end{figure*}

In addition to communication redundancy, flat communication over hierarchical networks in SpMM introduces another \mbox{communication} inefficiency. Modern GPU-accelerated HPC systems feature multi-level network architectures where intra-group connections 
significantly outperform inter-group links. For instance~\cite{tsubame4}, GPUs within the same node communicate via NVLink (450GB/s unidirectional link bandwidth on NVIDIA H100~\cite{nvidia-h100-web}), while cross-node communication relies on InfiniBand HDR (25GB/s~\cite{mellanox-hdr-whitepaper}), creating an 18$\times$ bandwidth disparity. However, standard SpMM implementations use flat all-to-all communication patterns that treat all process pairs equally, ignoring these bandwidth disparities. This topology-oblivious approach creates redundancy at group boundaries, as illustrated in~\Cref{fig/methods/motivation_hierarchical_comm}: (a) identical $\mathbf{B}$ rows ($B0, B1, B2$) sent to multiple destinations traverse slow inter-group links separately rather than being transmitted once and distributed within the destination group, and (b) partial $\mathbf{C}$ results ($C00, C10, C30$) targeting the same $\mathbf{C}$ row ($C0''$) cross group boundaries independently instead of being pre-aggregated within the source group. Such hierarchical redundancy significantly increases inter-group traffic, creating bottlenecks as system scale grows.

\subsection{Summary}\label{sec:mot_summary}
{Optimal performance requires jointly addressing both sources of inefficiency that current approaches lack: {sparsity-aware} communication strategy reduces communication volume but ignores network topology, while hierarchy-aware methods improve bandwidth utilization but retain strategy-level redundancy. }


\section{Overview}



We present \method{}, a unified optimization distributed SpMM framework that jointly addresses communication strategy redundancy and hierarchical network redundancy. \method{} minimizes total communication volume and inter-group communication through:

\begin{itemize}[leftmargin=1em]
    \item \textbf{Joint row-column communication strategy (\Cref{sec:joint}):} Exploits sparsity in row and column dimensions simultaneously, combining row-based and column-based patterns to minimize overall communication volume.
    
    \item \textbf{Hierarchical communication strategy (\Cref{sec:hierarchical}):} Adapts this joint row-column communication strategy to the network hierarchy {by separating communication into complementary stages} and scheduling them to minimize inter-group traffic while maximizing link utilization.
\end{itemize}

\Cref{fig/methods/overview} illustrates the overview of \method{}'s optimization {strategies}. Joint row-column optimization analyzes the sparse matrix structure to identify the optimal combination of row-based and column-based strategies, producing a unified communication plan that minimizes total communication volume beyond single-strategy limitations (\Cref{sec:joint}). The hierarchical communication strategy (\Cref{sec:hierarchical}) then separates the joint communication strategy into independent row-based and column-based communication, enabling each to eliminate inter-group redundancy individually through communication deduplication {(\Cref{subsec:redundancy_elimination})}. 
After independent optimization, these operations are recombined by exploiting their complementary network link requirements, achieving overlapped execution to maximize link utilization {(\Cref{subsec:overlapped_execution})}.

\section{Joint Row-Column Communication Strategy}\label{sec:joint}
This section presents our joint row-column optimization approach that eliminates communication redundancy by optimally assigning each nonzero element in the sparse matrix to use the row-based or column-based strategy. 
{This involves partitioning nonzeros based on either their row or column indices, whichever is communication optimal, then using the respective row- and column-based communication strategies for each set of nonzeros.}
We first introduce the detailed workflow of our method {(\Cref{subsec:prepos_workflow})}, then formulate the assignment decision as a covering optimization problem to minimize communication volume {(\Cref{subsec:prepos_formulation})}. We describe our solution algorithm {(\Cref{subsec:prepos_solve})}. Finally, we analyze the theoretical benefits under different sparsity structures (\Cref{subsec:analysis}).

\subsection{Workflow}\label{subsec:prepos_workflow}

\Cref{fig/methods/overview} shows the workflow of five stages of our joint row-column communication strategy:
\textbf{\circled{1} Matrix Sparsity Analysis:} {Each process $p$ analyzes the sparsity patterns of each off-diagonal submatrix $\mathbf{A}^{(p,q)}$ that determines data transfer from process $q$ to process $p$ for subsequent optimization-based strategy selection.}
\textbf{\circled{2} Communication Strategy Selection:} Each process $p$ solves a covering optimization problem to determine, for each nonzero in $\mathbf{A}^{(p,q)}$, whether to use row-based or column-based communication. Based on these decisions, $\mathbf{A}^{(p,q)}$ is partitioned into row-based and column-based components. Process $p$ retains the column-based portion and transfers the row-based portion to remote process $q$. Note that steps \circled{1} and \circled{2} are performed offline as a preprocessing phase and can be reused across multiple SpMM operations with the same sparsity pattern. 
\textbf{\circled{3} Remote Computation:} Process $q$ computes partial $\mathbf{C}$ results using the received row-based matrix portion and prepares required $\mathbf{B}$ rows based on the column indices of process $p$'s column-based portion. Both results are packed for transmission. 
\textbf{\circled{4} Communication:} Process $q$ transmits the computed partial $\mathbf{C}$ results and required $\mathbf{B}$ rows to process $p$. 
\textbf{\circled{5} Result Aggregation:} {Process $p$ combines the local computation results from received $\mathbf{B}$ rows and partial $\mathbf{C}$ results to produce its final result $\mathbf{C}^{(p,:)}$.}

\subsection{Problem Formulation for Optimal Communication}\label{subsec:prepos_formulation}
As outlined in the workflow, the second stage involves solving an optimization problem to determine the optimal communication strategy for each nonzero element. The key insight underlying this optimization is that each nonzero entry $(i,j)$ in the off-diagonal submatrices of $\mathbf{A}$ can be {covered} through either of two communication strategies: {communicating} the corresponding row of $\mathbf{B}$ (column-based strategy using column index $j$) or computing and {communicating} the corresponding row of $\mathbf{C}$ (row-based strategy using row index $i$). Since multiple nonzero entries may share the same row or column indices, communication of a single {$\mathbf{B}$ or $\mathbf{C}$} row can cover multiple entries simultaneously. For example, as illustrated in~\Cref{fig/method:complicated_pre_post_graph}, nonzero entries $\{b,c,d\}$ can be covered by communicating row $\mathbf{C}^{(0,:)}$ (row index 0), while entries $\{c,f,h\}$ can be covered by communicating row $\mathbf{B}^{(6,:)}$ (column index 6). The goal is therefore to \textbf{select a minimum-cost set of $\mathbf{B}$-row and $\mathbf{C}$-row communications that collectively cover all nonzero entries}.

A naive solution to this problem is to employ a greedy algorithm: count the number of nonzero entries covered by each row and column, sort them in descending order by coverage, and iteratively select rows or columns until all nonzero entries are covered. However, this greedy approach has two drawbacks: (1) it cannot guarantee global optimality, as greedy selections may lead to suboptimal overall communication costs, and (2) it incurs substantial overhead from iteratively counting coverage and updating statistics.

To address these limitations and achieve global optimality, we formalize the optimal strategy selection as a minimum weighted set cover problem, where the objective is to find the minimum-cost combination of $\mathbf{B}$-row and $\mathbf{C}$-row communications that covers all nonzero entries. Let $a_{ij} = 1$ if entry $(i,j)$ in the off-diagonal submatrices of $\mathbf{A}$ is nonzero, and $a_{ij} = 0$ otherwise. 
{We introduce binary decision variables to indicate whether each row is selected for communication. Since communicating different rows may incur different costs due to varying data volumes and network paths, we associate each variable with a cost coefficient. Specifically, for column index $j$ in the off-diagonal submatrices of $\mathbf{A}$ (rows of $\mathbf{B}$):}
\begin{equation}\footnotesize
x_j = \begin{cases}
1, & \text{if row } \mathbf{b}_{j,:} \text{ is communicated}, \\
0, & \text{otherwise}.
\end{cases}
\end{equation}
{with communication cost $w^{\mathrm{col}}_j$. For row index $i$ (rows of $\mathbf{C}$):}
\begin{equation}\footnotesize
y_i = \begin{cases}
1, & \text{if row } \mathbf{c}_{i,:} \text{ is communicated}, \\
0, & \text{otherwise},
\end{cases}
\end{equation}
{with communication cost $w^{\mathrm{row}}_i$.}

\noindent\textbf{Optimization Model.} The joint communication problem is formulated as a minimum weighted set cover problem:
\begin{equation}\label{eq:constraint}
\footnotesize
\begin{aligned}
\min \quad & \sum_{j} w^{\mathrm{col}}_j x_j + \sum_{i} w^{\mathrm{row}}_i y_i \\
\text{s.t.} \quad & x_j + y_i \geq a_{ij}, \quad \forall (i,j) \\
& x_j, y_i \in \{0,1\}, \quad \forall i,j
\end{aligned}
\end{equation}
where constraint in \eqref{eq:constraint} ensures that each nonzero entry $(i,j)$ in the off-diagonal submatrices of $\mathbf{A}$ is covered by communicating either row $\mathbf{b}_{j,:}$ (for column index) or row $\mathbf{c}_{i,:}$ (for row index).
\subsection{Optimal Solution via Minimum Weighted Vertex Cover}\label{subsec:prepos_solve}



\subsubsection{Formulation as a Graph Problem}

{The optimization problem essentially concerns row-column relationships established through nonzero entries in the off-diagonal submatrices of $\mathbf{A}$, which naturally admits a bipartite graph formulation. To solve this problem, we construct a bipartite graph $G = (\mathcal{R} \cup \mathcal{C}, E)$ where the left vertex set $\mathcal{R}$ and right vertex set $\mathcal{C}$ represent the row and column indices of the off-diagonal submatrices of $\mathbf{A}$, respectively, and the edge set $E = \{(i,j) \mid a_{ij} = 1, i \in \mathcal{R}, j \in \mathcal{C}\}$ connects row and column vertices for each nonzero entry in the off-diagonal submatrices of $\mathbf{A}$. Under this construction, the decision variables map directly to vertex selection: $y_i = 1$ selects row vertex $i \in \mathcal{R}$ (communicating $\mathbf{c}_{i,:}$), and $x_j = 1$ selects column vertex $j \in \mathcal{C}$ (communicating $\mathbf{b}_{j,:}$). Each vertex is weighted by its communication cost: $w^{\mathrm{row}}_i$ for row vertices and $w^{\mathrm{col}}_j$ for column vertices. Since covering all nonzero entries requires selecting at least one row or column for each entry, which corresponds to selecting at least one endpoint of every edge, the minimum weighted set cover problem reduces to the minimum weighted vertex cover problem on a bipartite graph, i.e., finding a minimum-weight vertex subset such that every edge is incident to at least one selected vertex:}


{
{\footnotesize
\begin{gather}
\min_{U \subseteq \mathcal{R} \cup \mathcal{C}} \sum_{i \in U \cap \mathcal{R}} w_i^{\mathrm{row}} + \sum_{j \in U \cap \mathcal{C}} w_j^{\mathrm{col}} \\
\text{s.t.} \quad i \in U \lor j \in U, \quad \forall (i,j) \in E
\end{gather}}
which directly corresponds to minimizing the total communication cost with constraints in our original formulation (Eqn.~\ref{eq:constraint}).
}

\begin{figure}[t!]
    \centering
    \includegraphics[clip,width=0.5\textwidth]{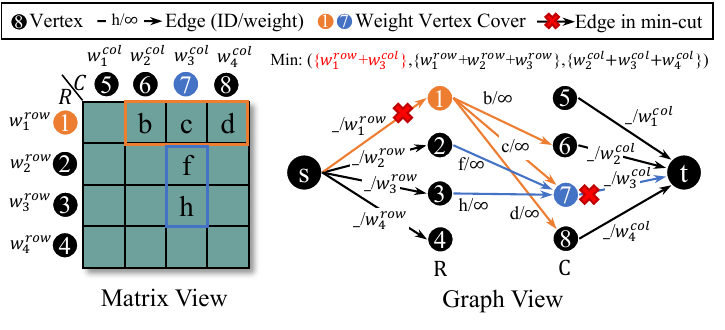}
    \vspace{-25pt}
    \caption{Optimal communication volume via minimum weighted vertex cover. The matrix view (left) shows nonzero entries, while the graph view (right) models the problem as a flow network. Matrix nonzeros correspond to graph edges {between $\mathcal{R}$ and $\mathcal{C}$}, while matrix rows and columns correspond to vertex sets $\mathcal{R}$ and $\mathcal{C}$, respectively. The minimum s-t cut (red crosses) yields the optimal solution: communicate the row of $\mathbf{C}$ indexed by row 1 and the row of $\mathbf{B}$ indexed by column 7 to cover all nonzero entries \{$b, c, d, f, h$\}.}
    \vspace{-15pt}
    \label{fig/methods/weighted_MVC}
\end{figure}
\subsubsection{Optimal Solution via Max-Flow Min-Cut}
To solve this vertex cover problem optimally, we further reduce it to a minimum $s$-$t$ cut problem by constructing a flow network. This reduction exploits the well-established correspondence between minimum weighted vertex cover and minimum cut in bipartite graphs~\cite{schrijver2003combinatorial}. We build the flow network by adding source $s$ and sink $t$, connecting $s$ to each row vertex $i \in \mathcal{R}$ with capacity $w^{\mathrm{row}}_i$, connecting each column vertex $j \in \mathcal{C}$ to $t$ with capacity $w^{\mathrm{col}}_j$, and assigning infinite capacity to all bipartite edges $(i,j) \in E$. 
{Since the minimum $s$-$t$ cut must avoid infinite-capacity edges, it can only cut edges incident to $s$ or $t$. Cutting edge $(s, i)$ corresponds to selecting row vertex $i$, and cutting edge $(j, t)$ corresponds to selecting column vertex $j$, together forming the optimal vertex cover.}

\Cref{fig/methods/weighted_MVC} demonstrates this approach on an example matrix. The minimum cuts occur at $s \to 1$ and $7 \to t$, yielding the optimal vertex cover $\{1, 7\}$, which corresponds to communicating the row of $\mathbf{C}$ indexed by row 1 and the row of $\mathbf{B}$ indexed by column 7 to cover all nonzero entries. The max-flow min-cut theorem~\cite{ford1956maximal} guarantees that this solution achieves the minimum communication cost, since the cut value directly corresponds to the total communication volume in our original problem. 

This problem can be solved efficiently using the standard max-flow algorithm, \textit{Dinic's algorithm}~\cite{dinic1970algorithm}, achieving $O(|V|^2|E|)$ time complexity where $|V| = |\mathcal{R}| + |\mathcal{C}|$ and $|E|$ represents the number of edges in the flow network. Importantly, this optimization is performed offline: as long as the sparsity pattern of matrix $\mathbf{A}$ remains unchanged, the computed communication strategy can be reused across multiple SpMM operations, effectively amortizing the overhead of solving the problem. {Moreover, since the optimization for each off-diagonal submatrix $\mathbf{A}^{(p,q)}$ is independent, these subproblems can be solved in parallel, further reducing overhead.}

\subsection{Theoretical Analysis of Communication Benefits}
\label{subsec:analysis}
In this subsection, we analyze the theoretical performance benefits of our joint optimization approach compared to existing methods. 
Here, we consider the common case where the communication costs are uniform for $\mathbf{B}$ rows and $\mathbf{C}$ rows. 

\subsubsection{Dominance Analysis}
For any off-diagonal submatrix $\mathbf{A}^{(p,q)}$, our joint optimization approach achieves communication volume 
\begin{equation} \footnotesize
V^{q,p}_{\text{joint}} = \mu \cdot N \cdot \mathit{sz}_{\mathrm{dt}}
\end{equation}
where $\mu$ represents the size of the minimum vertex cover obtained from our bipartite graph optimization. We can quantify the relative communication reductions compared to baseline row-based and column-based strategies as 
\begin{equation}\footnotesize
Red_{\text{col}} = 1 - \frac{\mu}{\bigl|\mathrm{Cols}(\mathbf{A}^{(p,q)})\bigr|}
\qquad
Red_{\text{row}} = 1 - \frac{\mu}{\bigl|\mathrm{Rows}(\mathbf{A}^{(p,q)})\bigr|}
\end{equation}


\subsubsection{Communication Reduction by Sparsity Patterns}

\begin{figure}[t!]
    \centering
    \includegraphics[clip,width=0.45\textwidth]{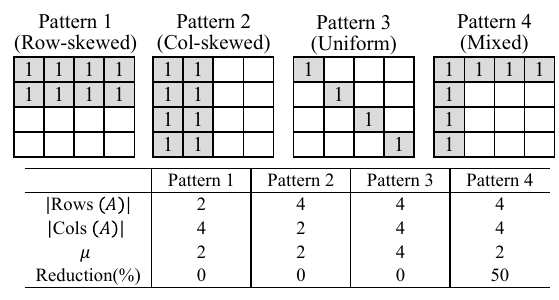}
    \vspace{-5pt}
    \caption{Different sparsity patterns and communication volume reduction. }
    \vspace{-10pt}
    \label{fig/methods/sparsity_pattern}
\end{figure}

The minimum vertex cover size $\mu$ depends on how high-degree vertices distribute across row and column {vertex sets (partitions)} in the bipartite graph. This distribution pattern determines the benefit.

\begin{figure*}[t!]
    \centering
    \includegraphics[clip,width=1\textwidth]{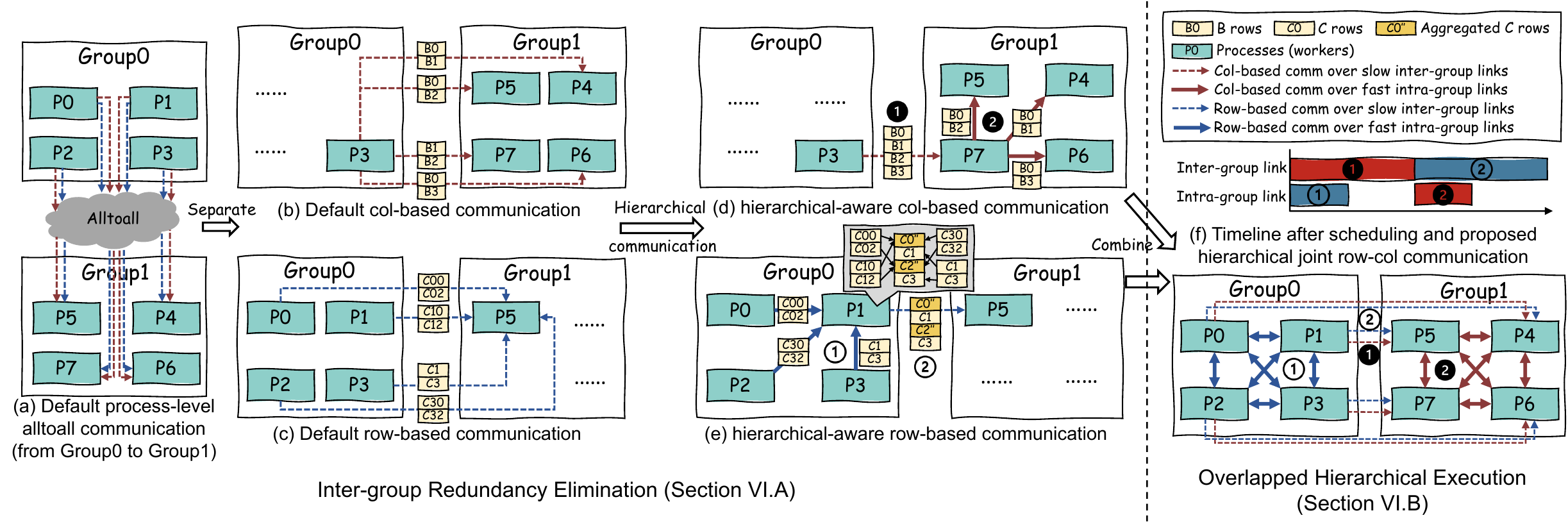}
    \vspace{-15pt}
    \caption{Overview of hierarchical communication strategy. For clarity, only Group0-to-Group1 transfers are shown (reverse direction omitted), and row/column-based communication illustrates operations for a single process (others operate similarly). }
    \vspace{-7pt}
\label{fig/method:hierarchical_comm}
\end{figure*}

\textbf{Low-reduction scenarios} occur when 
\begin{equation}\footnotesize
\mu \approx \min\left\{\bigl|\mathrm{Cols}(\mathbf{A}^{(p,q)})\bigr|, \bigl|\mathrm{Rows}(\mathbf{A}^{(p,q)})\bigr|\right\}
\end{equation}
\Cref{fig/methods/sparsity_pattern} illustrates two such cases: (1) skewed patterns where high-degree vertices cluster in one partition (Patterns 1--2), and (2) uniform patterns where all vertices have similar low degrees (Pattern 3). Both cases require the vertex cover to include most vertices on one partition, limiting joint optimization benefits.

\textbf{High-reduction scenarios} occur when 
\begin{equation}\footnotesize
\mu \ll \min\left\{\bigl|\mathrm{Cols}(\mathbf{A}^{(p,q)})\bigr|, \bigl|\mathrm{Rows}(\mathbf{A}^{(p,q)})\bigr|\right\}
\end{equation}
This requires high-degree vertices distributed across both partitions, enabling efficient bipartite coverage. Pattern 4 in 
\Cref{fig/methods/sparsity_pattern} demonstrates this: high-degree 
vertices on both sides create complementary coverage, achieving 50\% reduction compared to single-strategy approaches where only one partition is {selected.}

{In summary, the reduction potential is determined by the sparsity pattern characteristics. When high-degree vertices are well-distributed across both partitions, the joint strategy achieves significant communication savings; conversely, skewed or uniform patterns limit the optimization gains. Nonetheless, the joint strategy guarantees no performance degradation as it generalizes both single strategies as special cases.}

\section{Hierarchical Communication Strategy}
\label{sec:hierarchical}
{This section presents how we adapt the joint row-column communication strategy to hierarchical networks, where intra-group links offer significantly higher bandwidth than inter-group ones.}
As discussed in~\Cref{sec:motivation}, flat communication patterns create redundant inter-group transfers. To address this, we separate the joint strategy into \textit{row-based} and \textit{column-based} communication, partition them into \textit{fine-grained stages} to eliminate inter-group redundancy (\Cref{subsec:redundancy_elimination}), and recombine them through \textit{complementary scheduling} to fully utilize the bandwidth of different tier links (\Cref{subsec:overlapped_execution}).

\subsection{Inter-group Redundancy Elimination}
\label{subsec:redundancy_elimination}
\subsubsection{Communication Pattern Separation}
The joint optimization result combines \textit{row-based} and \textit{column-based} communications. However, these exhibit different redundancy patterns in hierarchical networks. {Specifically, }\textit{column-based} communication shows multiple processes fetching the same $\mathbf{B}$ rows across group boundaries (e.g., $B0, B1, B2$ transmitted multiple times from Group0 to Group1 in \Cref{fig/method:hierarchical_comm}(b)), while \textit{row-based} communication has processes independently sending partial $\mathbf{C}$ results that could be pre-aggregated (e.g., $C00, C10, C30$ rows sent separately in \Cref{fig/method:hierarchical_comm}(c)). These distinct redundancy patterns motivate our approach: we separate the joint communication plan into \textit{row-based} and \textit{column-based} operations, enabling targeted optimization for each pattern.

\subsubsection{Hierarchical Aggregation Strategies}\label{sec:subsubsection_prepose_redundancy}

We eliminate inter-group redundancy through hierarchy-aware aggregation strategies specific to {\textit{row-based} and \textit{column-based} communication respectively:}

\noindent\textbf{\textit{Column-based} redundancy elimination.} For matrix $\mathbf{B}$ distribution, we adopt a three-step method~\cite{zhang2025cola}: {(1)} source group aggregation, {(2)} inter-group transfer, and {(3)} intra-group distribution. {Specifically,}
process $P_3$ in Group0 first collects all its $\mathbf{B}$ rows required by Group1 processes ($P_4$--$P_7$), obtaining the deduplicated set \{$B_0$, $B_1$, $B_2$, $B_3$\} ({Step 1}). It then performs a single \textit{inter-group} transfer ({Step 2, i.e.,~\Cref{fig/method:hierarchical_comm}(d) Stage \circled{1}}) to Group1's representative, which distributes the rows internally using fast \textit{intra-group} links ({Step 3, i.e.,~\Cref{fig/method:hierarchical_comm}(d) Stage \circled{2}}). This reduces inter-group traffic from 8 to 4 row transfers, ensuring each unique $\mathbf{B}$ row crosses group boundaries only once.

\noindent\textbf{\textit{Row-based} redundancy elimination.} For matrix $\mathbf{C}$ partial results, we propose a two-stage hierarchical aggregation: (1) \textit{intra-group} pre-aggregation of partial results, and (2) \textit{inter-group} transmission of aggregated data. In the first stage, a representative process within each source group collects partial $\mathbf{C}$ results from group members and aggregates those destined for the same final $\mathbf{C}$ rows ({\Cref{fig/method:hierarchical_comm}(e) Stage \circledwhite{1}}). \Cref{fig/method:hierarchical_comm}(e) shows process $P_1$ in Group0 collects partial results from $P_0$, $P_2$, and $P_3$, then sums partial results targeting the same output $\mathbf{C}$ row (e.g., aggregating $C_{00}$, $C_{10}$, $C_{30}$ into $C_0''$). This pre-aggregation uses fast \textit{intra-group} links and reduces the data volume before crossing group boundaries. In the second stage, the representative transmits only the aggregated results to target processes in Group1 ({\Cref{fig/method:hierarchical_comm}(e) Stage \circledwhite{2}}). This approach reduces \textit{inter-group} transfers from 8 partial $\mathbf{C}$ results to 4 aggregated $\mathbf{C}$ results, decreasing traffic over slow inter-group links.

\subsection{{Overlapping Complementary Stages
}}
\label{subsec:overlapped_execution}
{In this section, we design an overlapping scheduling strategy for row- and column-based communication (\Cref{subsubsec:complementary}) and present its implementation workflow (\Cref{subsubsec:workflow}).}
\subsubsection{Scheduling Complementary Stages}
\label{subsubsec:complementary}
\input{figures/algorithm1}

The row-based and column-based redundancy elimination (\Cref{sec:subsubsection_prepose_redundancy}) exhibit natural complementarity: when one uses intra-group links, the other uses inter-group links. This complementary execution order is determined by the source of redundancy in each type. Row-based communication must aggregate partial $\mathbf{C}$ results within source groups before inter-group transmission (\Cref{fig/method:hierarchical_comm}(e)~\circledwhite{1}$\rightarrow$\circledwhite{2}), while column-based communication must fetch $\mathbf{B}$ rows across groups before intra-group distribution in destination groups (\Cref{fig/method:hierarchical_comm}(d)~\circled{1}$\rightarrow$\circled{2}).

Since no dependencies exist between row- and column-based communications, we exploit this complementary network usage through a two-stage overlapping strategy (\Cref{fig/method:hierarchical_comm}(f)).
Stage I pairs row-based intra-group aggregation (\Cref{fig/method:hierarchical_comm}(f)~\circledwhite{1}) with column-based inter-group fetching (\Cref{fig/method:hierarchical_comm}(f)~\circled{1}), allowing source groups to aggregate partial $\mathbf{C}$ results with intra-group links while destination groups simultaneously fetch required $\mathbf{B}$ rows with inter-group links. Stage II reverses this pattern: row-based inter-group transmission (\Cref{fig/method:hierarchical_comm}(f)~\circledwhite{2}) occurs alongside column-based intra-group distribution (\Cref{fig/method:hierarchical_comm}(f)~\circled{2}), ensuring aggregated $\mathbf{C}$ results cross group boundaries while fetched $\mathbf{B}$ rows are distributed within destination groups. This complementary execution maintains continuous utilization of both network tiers without contention.
\begin{figure*}[ht!]
    \centering
    \includegraphics[clip,width=1\textwidth]{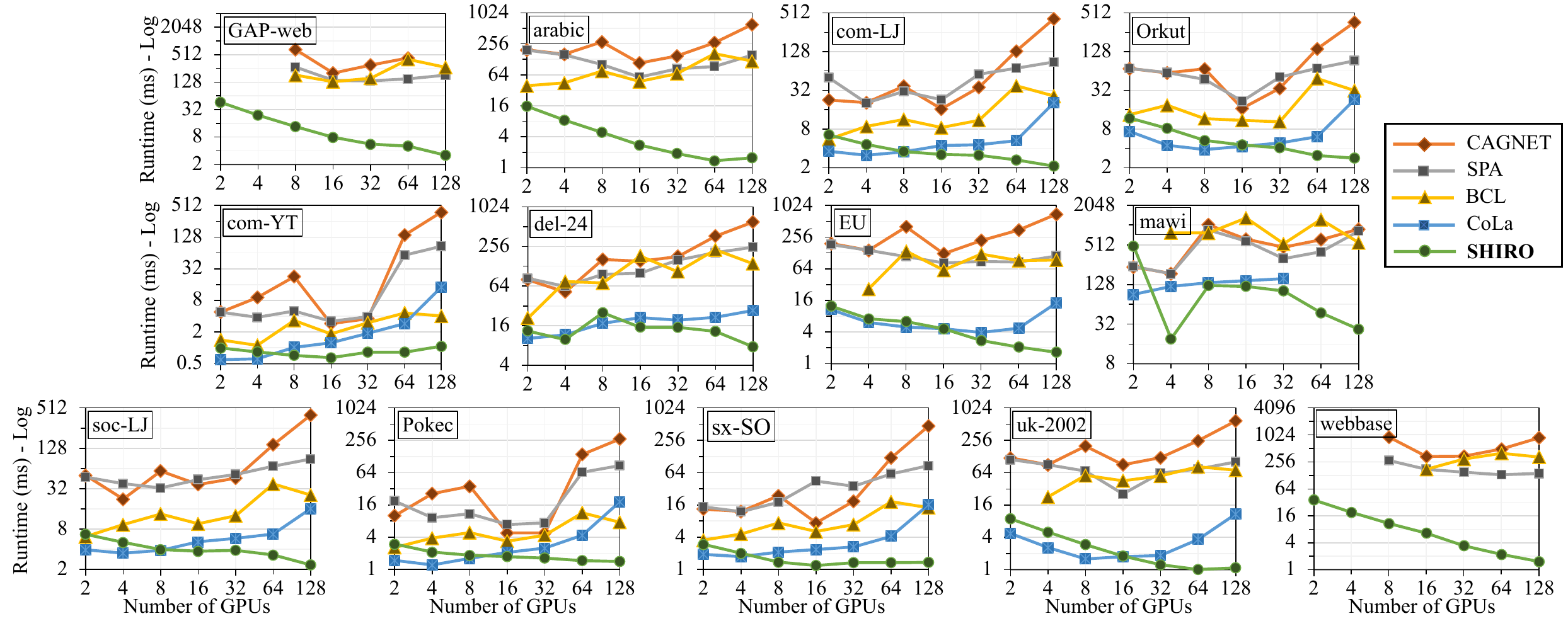}
    \vspace{-15pt}
    \caption{Runtime comparison ({\bf{Y-axis log-scale}}) with baselines and strong scaling for different datasets. Due to out-of-memory (OOM) and runtime errors, some data could not be collected. The dense column number ($N$) is 32.} 
    \vspace{-10pt}
    \label{fig/experiment:scaling}
\end{figure*}

\subsubsection{Communication Workflow}
\label{subsubsec:workflow}
\Cref{fig/method:hierarchical_comm}(f) and~\Cref{alg:hierarchical} present the complete hierarchical communication from a per-process perspective. The algorithm takes local sparse matrix blocks $\mathbf{A}_{row}$ and $\mathbf{A}_{col}$, dense matrix $\mathbf{B}$, and group ID as inputs, executing two stages of parallel row-column communications. Stage I combines column-based inter-group B fetching (lines 1--4, stage~I.\circled{1}), where processes collect and send required $\mathbf{B}$ rows to peers in other groups, with row-based intra-group C aggregation (lines 5--7, stage~I.\circledwhite{1}), where processes compute SpMM with $\mathbf{A}_{row}$ then exchange partial $\mathbf{C}$ results within groups. Stage II then pairs row-based inter-group $\mathbf{C}$ transmission (lines 8--11, stage~II.\circledwhite{2}), where processes aggregate the partial results from stage~I.\circledwhite{1} destined for the same remote peer before transmission, with column-based intra-group $\mathbf{B}$ distribution (lines 12--14, stage~II.\circled{2}), where processes redistribute the $\mathbf{B}$ rows received in stage~I.\circled{1} within their groups. The algorithm synchronizes both communication stages (lines 13 and 15) and returns the received results: $B_{dist}$ from column-based communication and $C_{recv}$ from row-based communication {for subsequent computations.}




\begin{table}[t]
\centering
\vspace{5pt}
\caption{Sparse matrices datasets. All matrices are square (\#rows = \#cols). \textit{NNZ} denotes the number of nonzeros.}
\label{tab:dataset}
\vspace{-5pt}
\resizebox{\linewidth}{!}{%
\begin{tabular}{lrrrr}
\toprule
\textbf{Matrix} & \textbf{\#Rows} & \textbf{NNZ} & \textbf{Density} & \textbf{Domain} \\
\midrule
com-Youtube (com-YT)     & 1.1M   & 6.0M   & 4.64e-06 & Social \\
soc-Pokec (Pokec)        & 1.6M   & 30.6M  & 1.15e-05 & Social \\
sx-stackoverflow (sx-SO) & 2.6M   & 36.2M  & 5.35e-06 & Q\&A \\
soc-LiveJournal (soc-LJ) & 4.8M   & 69.0M  & 2.94e-06 & Social \\
com-LiveJournal (com-LJ) & 4.0M   & 69.4M  & 4.34e-06 & Social \\
delaunay\_n24 (del24)    & 16.8M  & 100.7M & 3.58e-07 & Mesh \\
europe\_osm (EU)         & 50.9M  & 108.1M & 4.17e-08 & Road \\
mawi\_69M (mawi)         & 68.9M  & 143.4M & 3.02e-08 & Traffic \\
com-Orkut (Orkut)        & 3.1M   & 234.4M & 2.48e-05 & Social \\
uk-2002 (uk-2002)        & 18.5M  & 298.1M & 8.69e-07 & Web \\
arabic-2005 (arabic)     & 22.7M  & 640.0M & 1.24e-06 & Web \\
webbase-2001 (webbase)   & 118.1M & 1.02B  & 7.31e-08 & Web \\
GAP-web (GAP-web)        & 50.6M  & 1.93B  & 7.53e-07 & Web \\
\midrule
OGB-mag240M (Mag240M)    & 121.7M & 2.59B  & 1.75e-07 & GNN \\
OGB-papers100M (Papers)  & 111.1M & 3.23B  & 2.62e-07 & GNN \\
IGB260M (IGB260M)        & 269.3M & 3.72B  & 5.13e-08 & GNN \\
\bottomrule
\end{tabular}
}
\vspace{-10pt}
\end{table}

\section{Evaluation}
\subsection{Experimental Setup}
\subsubsection{Design}
Our experiments consist of five parts: (1) SpMM performance and strong scaling comparison across various sparse matrices; (2) ablation studies to validate each optimization; (3) experiments on varying dense matrix columns to demonstrate generality; (4) a GNN training case study to contextualize preprocessing overhead and evaluate end-to-end performance on large-scale matrices; and (5) a portability study to demonstrate that our communication optimizations generalize beyond NVIDIA-based systems. Experiments (1)--(4) are conducted on NVIDIA GPUs, while (5) is conducted on an Intel-based platform.

\subsubsection{Hardware} 
Experiments (1)--(4) are conducted on the TSUBAME4.0 supercomputer~\cite{tsubame4}, and the portability study (5) is conducted on the Aurora supercomputer~\cite{allcock2025auroraarchitectingargonnesexascale}. Each TSUBAME4.0 node is equipped with dual AMD EPYC 9654 processors and 4 NVIDIA H100 SXM5 GPUs with 94GB HBM2e memory, interconnected via NVLink 4.0, providing 450GB/s unidirectional bandwidth per GPU. The nodes are interconnected via InfiniBand NDR200 (25\,GB/s per link) in a fat-tree topology. Each Aurora node consists of dual Intel\textsuperscript{\textregistered} Xeon\textsuperscript{\textregistered} CPU Max Series processors and 6 Intel\textsuperscript{\textregistered} Data Center GPU Max 1550 GPUs; each GPU contains two Ponte Vecchio (PVC) tiles, yielding 12 tiles per node connected via Xe Link at 15GB/s per link. Nodes are interconnected via the Cray Slingshot-11 network at 200GB/s per node.

\subsubsection{Datasets}
We use diverse sparse matrices from the SuiteSparse Matrix Collection~\cite{davis2011university} (\Cref{tab:dataset}) that span multiple domains with varying sparsity patterns, and are widely used in recent related work~\cite{zhang2025cola, block2024two, abubaker2024spcomm3d}. {For the GNN case study, we select three large-scale sparse matrices from widely used GNN benchmarks~\cite{hu2020ogb, hu2021ogb, igbdatasets}.}


\subsubsection{Implementation}
We represent sparse matrices in the CSR and COO formats.
For single-node SpMM, we use {PyTorch (v2.5.1)}'s \texttt{torch.sparse.mm}~\cite{paszke2019pytorch}, which calls CSR version's SpMM in cuSPARSE~\cite{cusparse}.
Multi-node communication is implemented using PyTorch distributed~\cite{pytorch_distributed} with NCCL (v2.21.5)~\cite{nccl} and XCCL~\cite{oneccl} backends, utilizing \texttt{alltoall} primitive.
For the weighted minimum vertex cover problem on the bipartite graph, we implement Dinic's algorithm~\cite{dinic1970algorithm} in C++ via the standard max-flow reduction.
For the uniform-weight case, which reduces to unweighted minimum vertex cover, we provide a faster C++ implementation based on maximum bipartite matching and K\"{o}nig's theorem~\cite{schrijver2003combinatorial}. 

\subsubsection{{SpMM} Baselines}


To validate the effectiveness of our methods,
we select baselines sharing similar design choices with our approach except for key differences 
in: (1) matrix partitioning (1D, 1.5D, 2D); (2) sparsity awareness (oblivious vs. aware); and (3) hierarchy awareness (oblivious vs. aware).
Specifically, we compare against four state-of-the-art distributed SpMM methods: CAGNET~\cite{tripathy2020reducing} (1.5D/stationary A, sparsity-oblivious, NCCL~\cite{nccl}), SPA~\cite{mukhopadhyay2024sparsity} (1.5D/ stationary A, column-based sparsity-aware, NCCL~\cite{nccl}), BCL~\cite{brock2024rdma} (2D/stationary C, sparsity-oblivious, NVSHMEM~\cite{nvshmem}), and CoLa~\cite{zhang2025cola} (1D/stationary A, column-based sparsity-aware with hierarchical-awareness, NVSHMEM~\cite{nvshmem}). For CAGNET and SPA, we 
set the replication factor to 4, which delivers near-optimal performance in their original papers. Matrix (graph) reordering is disabled to ensure fair comparison and maintain applicability to diverse matrix types. 
We conduct 5 warm-up runs followed by 100 timed runs. We collect the end-to-end execution time of each run and report the average time over the 100 runs.

\begin{figure*}[t]
    \centering
    \subfigure[Comparison of total global communication volume between the default column-based strategy and joint row-column strategy.
    ]{
        \begin{minipage}[b]{0.47\textwidth}
            \includegraphics[width=1\textwidth]{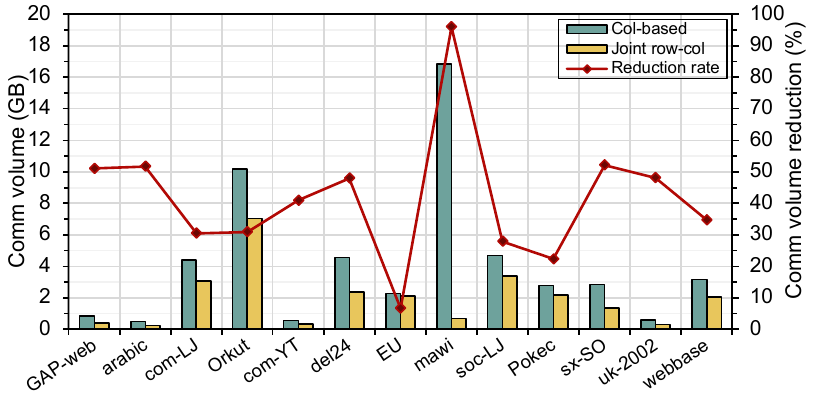}
            \vspace{-20pt}
        \end{minipage}
        \label{fig/experiment:comm_reduction_mvc}
    }
    \subfigure[Comparison of total inter-node communication volume between joint row-column strategy and the same strategy with hierarchical communication.
    ]{
        \begin{minipage}[b]{0.47\textwidth}
            \includegraphics[width=1\textwidth]{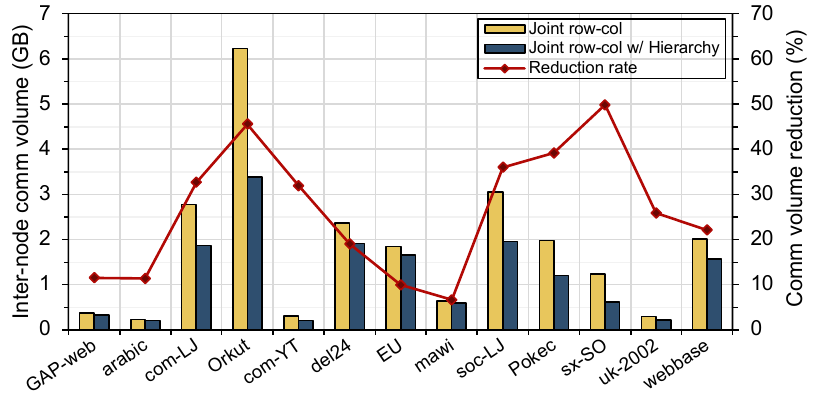}
            \vspace{-20pt}
        \end{minipage}
        \label{fig/experiment:comm_reduction_hierarchy}
    } 
    \vspace{-15pt}
    \caption{Global communication volume reduction by joint row-column communication and inter-node communication volume reduction by hierarchical communication strategy. nGPUs = 32.}
    \vspace{-5pt}
\end{figure*}

\begin{figure*}[t!]
    \centering
    \includegraphics[clip,width=\textwidth]{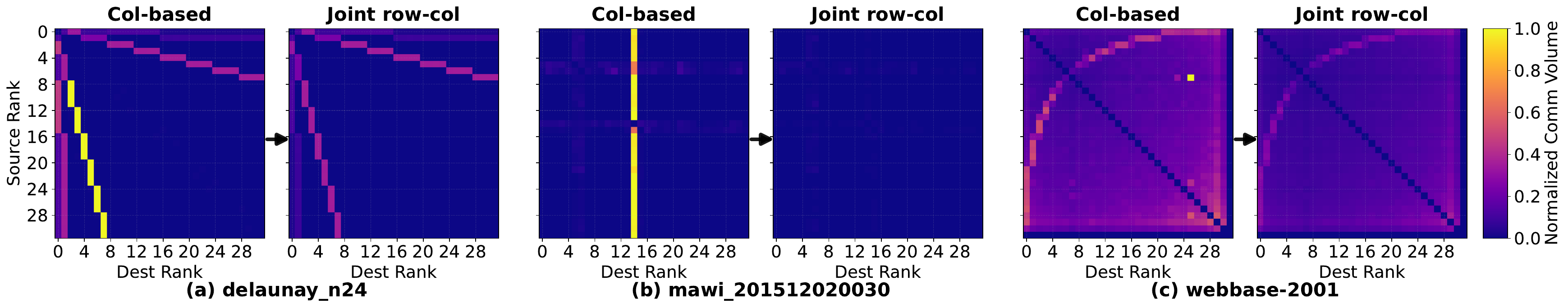}
    \vspace{-20pt}
    \caption{
Inter-process communication patterns before and after applying joint row-column strategy across datasets: achieving lower volume and better balance. Communication volumes are normalized by the maximum value within each respective dataset.}
    \vspace{-13pt}    
    \label{figures/experiment/comm_pattern}
\end{figure*}

\subsection{Overall Performance}
In this subsection, we compare our method with baselines across different datasets and GPU counts with 32 dense columns, as shown in \Cref{fig/experiment:scaling}. 
Some results are omitted due to out-of-memory (OOM) or runtime errors.
Compared to CAGNET and BCL, which need to fetch B blocks (CAGNET) or both A and B blocks (BCL) without considering sparsity patterns, our method consistently outperforms them across all GPU counts and datasets. Notably, CAGNET's poor performance stems from suboptimal cuSPARSE usage and synchronous broadcast-based communication that causes process idling and low utilization.
These results demonstrate the benefit of sparsity-aware communication strategies on diverse sparsity patterns. {Additionally, our method also outperforms SPA, which only employs column-based communication and lacks hierarchical communication.} 

Compared to CoLa, our method is slower when using 4 or fewer GPUs because each TSUBAME node contains 4 GPUs interconnected with all-to-all NVLink. At this scale, intra-node communication overhead is minimal and computation remains the bottleneck, where CoLa's computational optimizations and fine-grained RDMA communication for better computation-communication overlap provide better performance. However, when scaling to 8 or more GPUs across multiple nodes, communication becomes the primary bottleneck and our method outperforms CoLa on most datasets, with speedups increasing as GPU count grows.
This confirms that our communication optimizations effectively reduce communication overhead and improve overall SpMM performance at scale.

\subsection{Scaling}
\Cref{fig/experiment:scaling} demonstrates the strong scaling of our method compared to baselines across different datasets. The experiment starts from 2 GPUs to 128 GPUs. All baselines can only scale up to 8 GPUs (2 compute nodes), after which execution time increases with additional GPUs due to communication overhead. In contrast, our method exhibits decreasing execution time as GPU count grows on most datasets, scaling up to 128 H100 GPUs. Notably, our method achieves high scaling efficiency on datasets such as GAP-web, uk-2002, and webbase. One exception is the mawi dataset, where 4 GPUs achieve the best performance. Profiling with NVIDIA Nsight Systems~\cite{nsight-systems} reveals that: (1) our method reduces communication overhead significantly on this dataset (shown in \Cref{fig/experiment:comm_reduction_mvc}), shifting the bottleneck to computation; and (2) due to irregular sparsity, cuSPARSE launches the SpMM kernel with a grid dimension of (1,1,1) on all configurations except 4-GPU, resulting in poor computational performance on these configurations. Overall, this strong scaling result further validates that our method effectively reduces communication overhead, which limits scalability.

\subsection{Effectiveness of Proposed Methods}
This subsection validates the effectiveness of our proposed methods by examining the reduction in communication volume and overall runtime. We conduct all experiments on 32 GPUs with 64 dense columns. 
\subsubsection{Reduction in Communication Volumes}~\\
\noindent\textbf{Joint row-column sparsity-aware strategy.} We compare the total communication volume between the column-based strategy and our joint row-column strategy. \Cref{fig/experiment:comm_reduction_mvc} shows the total communication volume and the reduction ratio achieved by our method. The reduction ratio varies across different datasets, but our method consistently reduces communication volume on all datasets, validating its effectiveness for communication optimization. Notably, our method achieves the most significant reduction on the mawi dataset, eliminating 96\% of the communication volume compared to the default strategy. This communication reduction translates directly to performance gains, as mawi also exhibits the highest speedup (nearly 6$\times$) in the overall runtime results presented later. 

We also analyze the inter-process communication patterns between column-based and joint row-column communication strategies, as illustrated in~\Cref{figures/experiment/comm_pattern}.
Due to space limitations, we present three representative datasets that exhibit more imbalanced communication patterns compared to others.
In these heatmaps, the y-axis represents the source rank and the x-axis represents the destination rank, with brighter regions indicating higher communication volume between rank pairs. 
Across all three datasets, the joint row-column strategy not only reduces the overall communication volume but also significantly diminishes the communication between previously heavy-communicating rank pairs (eliminating the bright spots in the figures), resulting in a more balanced communication pattern. 
Notably, two of these datasets (delaunay\_n24 and mawi) correspond to symmetric matrices from undirected graphs, where the ideal communication pattern should also be symmetric. While the column-based strategy produces asymmetric and imbalanced patterns, our joint row-column strategy preserves the inherent symmetry of the communication.
A more balanced communication pattern typically translates to better performance for all-to-all communication.

\noindent\textbf{Hierarchical communication strategy.} 
Building on our joint row-column strategy, we compare the total inter-node communication volume with and without the hierarchical communication strategy, as shown in \Cref{fig/experiment:comm_reduction_hierarchy}.
Similarly, the reduction varies across matrices, but our hierarchical strategy consistently reduces inter-node communication on all datasets. The reduction is particularly pronounced on com-LJ, Orkut, Pokec, and sx-SO datasets, which correspondingly demonstrate substantial improvements in the overall runtime results.



\begin{figure}[t!]
    \centering
    \includegraphics[clip,width=0.48\textwidth]
    {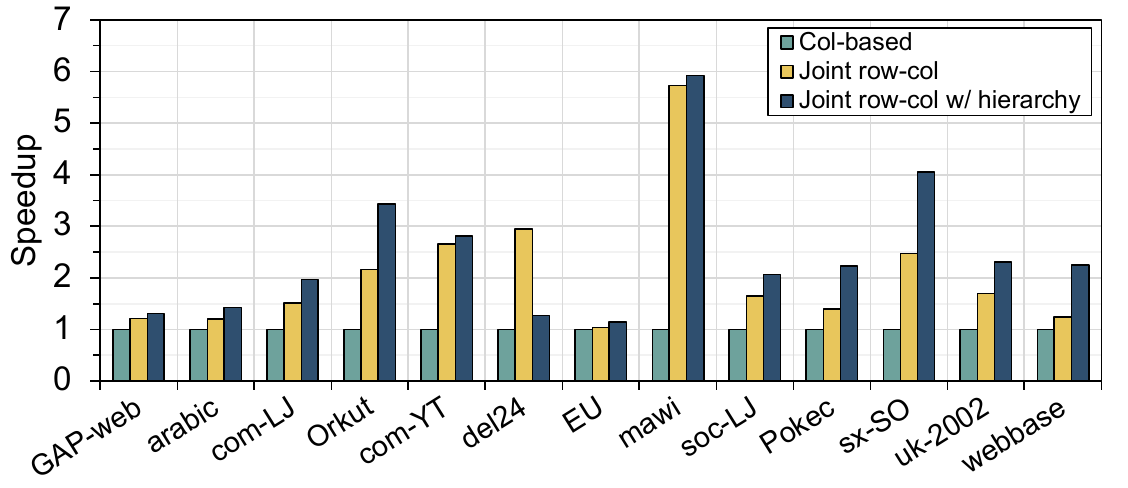}
    \vspace{-15pt}
    \caption{Step-wise optimization (runtime) results. nGPUs = 32.}
    \vspace{-10pt}
    \label{fig/experiment:stepwise}
\end{figure}

\subsubsection{Reduction in Overall Runtime} \Cref{fig/experiment:stepwise} presents ablation study results with 64 dense columns on 32 GPUs. The relative speedups of both methods vary across datasets, reflecting the diverse communication patterns inherent to different sparse matrix structures. Specifically, our joint row-column communication strategy consistently achieves speedups across all datasets, demonstrating its ability to reduce communication redundancy and improve overall performance. The hierarchical communication strategy delivers performance gains on most datasets, demonstrating that our approach reduces inter-node communication while maintaining high network link utilization, thereby improving both communication and overall performance. The only exception is del24, which exhibits highly imbalanced communication patterns. For this case, decomposing a collective communication into {several imbalanced communications} reduces network link utilization, thereby degrading performance. Overall, these results validate our communication analysis: datasets with greater communication reduction consistently achieve higher speedups across both optimization methods.

\begin{figure}[t!]
    \centering
    \includegraphics[clip,width=0.48\textwidth]{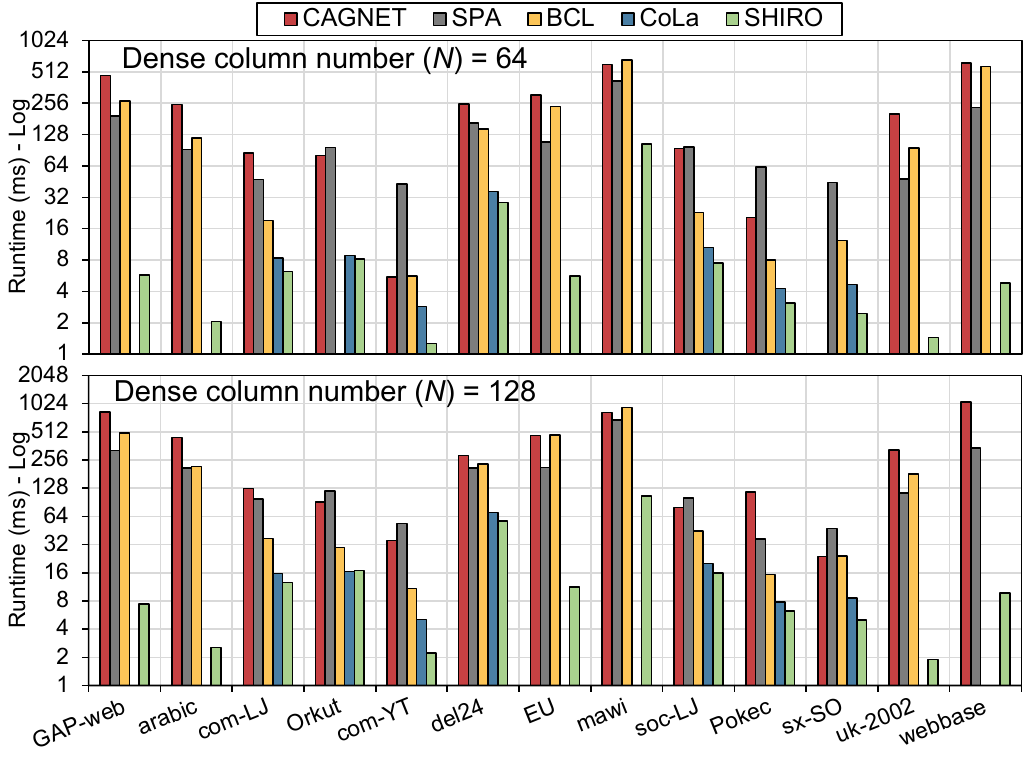}
    \vspace{-24pt}
    \caption{{Performance with varying dense column numbers ($N$ = 64, 128). Missing data indicate OOM. nGPUs = 32.}}
    \vspace{-5pt}
    \label{figures/experiment/performance_different_k}
\end{figure}
\subsection{{Performance of Different Dense Columns}}
{The number of dense columns ($N$) varies across applications, e.g., the feature dimension in GNNs. To assess the sensitivity of our method to this parameter, we evaluate performance with $N$ = 64 and 128, as shown in \Cref{figures/experiment/performance_different_k}. Some data points are missing due to OOM. Overall, our method outperforms other baselines on most datasets across different $N$ values, which validates the effectiveness of our approach for varying dense matrix sizes. Additionally, on most datasets, our method exhibits linear scaling with $N$, indicating that the execution is communication throughput-bound.
}

\begin{table}[t]
\centering
\caption{{Performance comparison in distributed GNN training (128 GPUs). \textit{Comm.} and \textit{Total} denote the communication and execution time of SpMM. \textit{Training (+ Prep.)} reports end-to-end training time and one-time preprocessing (\textit{Prep.}) overhead. \textit{Prep. Ratio} refers to the ratio of \textit{Prep.} to \textit{Training}.}}
\vspace{-5pt}
\label{tab:gnn}
\setlength{\tabcolsep}{3.5pt} 
\resizebox{0.9\linewidth}{!}{
\begin{tabular}{lc cc c c}
\toprule
\multirow{2}{*}{\textbf{Dataset}} & 
\multirow{2}{*}{\textbf{Methods}} & 
\multicolumn{2}{c}{\textbf{SpMM (s)}} & 
\multirow{2}{*}{\begin{tabular}[c]{@{}c@{}}\textbf{Training} \\ \textbf{(+ Prep.) (s)}\end{tabular}} & 
\multirow{2}{*}{\begin{tabular}[c]{@{}c@{}}\textbf{Prep.} \\ \textbf{Ratio$^*$}\end{tabular}} \\ 
\cmidrule(lr){3-4}
& & \textbf{Comm.} & \textbf{Total} & & \\
\midrule

\multirow{3}{*}{Papers}  
& BCL    & --    & 1009.4 & --    & -- \\
& PyG    & 525.1 & 544.9  & 673.7 & -- \\
& \textbf{SHIRO} & \textbf{324.2} & \textbf{335.2} & \textbf{377.3 (+38.7)} & \textbf{9.3\%} \\

\midrule

\multirow{3}{*}{Mag240M} 
& BCL    & --    & 786.2  & --    & -- \\
& PyG    & 305.2 & 325.5  & 349.4 & -- \\
& \textbf{SHIRO} & \textbf{255.2} & \textbf{263.2} & \textbf{299.1 (+45.7)} & \textbf{13.2\%} \\

\midrule

\multirow{3}{*}{IGB260M} 
& BCL    & --    & 1303.7 & --    & -- \\
& PyG    & 323.4 & 343.2  & 356.6 & -- \\
& \textbf{SHIRO} & \textbf{210.5} & \textbf{217.3} & \textbf{229.1 (+17.1)} & \textbf{6.9\%} \\

\bottomrule
\multicolumn{6}{l}{\scriptsize $^*$Ratio = Prep. / (Prep. + Training Time) $\times$ 100 $\%$.}
\vspace{-5pt}
\end{tabular}
}
\end{table}

\subsection{Case Study on GNNs}
{This subsection evaluates the end-to-end performance and preprocessing overhead of our method in large-scale full-batch GNN training.
We run GNN training on 128 NVIDIA H100 GPUs, with the preprocessing step (solving the Minimum Weighted Vertex Cover (MWVC) problem) using 256 MPI processes. Due to memory constraints, we set the feature size and hidden dimension to 128 (i.e., dense column number $N$ in SpMM is 128) for Papers and Mag240M, and 64 for IGB260M. Training is conducted over 200 epochs, resulting in 1000 SpMM calls. We implement GNN training based on PyTorch Geometric~\cite{fey2019fast} with default column-based SpMM (PyG) and \method{}. Since CoLa fails on matrices of this scale due to OOM, we use the second-best SpMM baseline, BCL, for SpMM performance reference. We omit BCL's communication time as its highly asynchronous design makes accurate measurement difficult.}

{\textbf{Performance.} We measure the communication time within SpMM, the total SpMM time, and the end-to-end training time (\Cref{tab:gnn}). The results reveal that SpMM constitutes the primary bottleneck in GNN training, with communication accounting for over 90\% of SpMM time. With an average SpMM time of hundreds of milliseconds per call, efficient large-scale distributed SpMM becomes essential. Compared to BCL's SpMM, \method{} achieves 3--6$\times$ speedup, demonstrating its effectiveness for SpMM on large-scale sparse matrices. Against column-based SpMM (PyG), \method{} reduces communication volume, lowering communication time by 1.20--1.62$\times$ and yielding 1.24--1.63$\times$ SpMM speedup. This translates to 1.17--1.79$\times$ speedup in end-to-end GNN training. Notably, on Papers, the end-to-end speedup exceeds the SpMM speedup because \method{} also mitigates load imbalance across processes, which is not captured in SpMM timing but reflected in epoch-end synchronization barriers. These results validate that \method{} effectively reduces communication overhead, thereby accelerating both large-scale SpMM and GNN training.}

{\textbf{Preprocessing Overhead.} The preprocessing (MWVC) time and its proportion of the total training time are shown in the last two columns of \Cref{tab:gnn}. This ratio varies across datasets (6.9\%--13.2\%) depending on the sparsity pattern, yet remains low enough to be amortized within a single training run. Moreover, compared to SpMM execution time, the preprocessing overhead is also modest (e.g., 17.1s vs. 217.3s on IGB260M), enabling amortization in workloads involving repeated SpMM with the same sparse matrix. This ratio would further decrease with larger feature sizes or hidden dimensions, as SpMM time grows while MWVC time remains constant. Additionally, the MWVC solution can be reused across subsequent runs as long as the sparse matrix remains unchanged, e.g., during hyperparameter tuning. Even when including the preprocessing time, \method{} still achieves better SpMM and end-to-end performance than the baselines.}

\begin{figure}[t!]
    \centering
    \includegraphics[clip,width=0.45\textwidth]
    {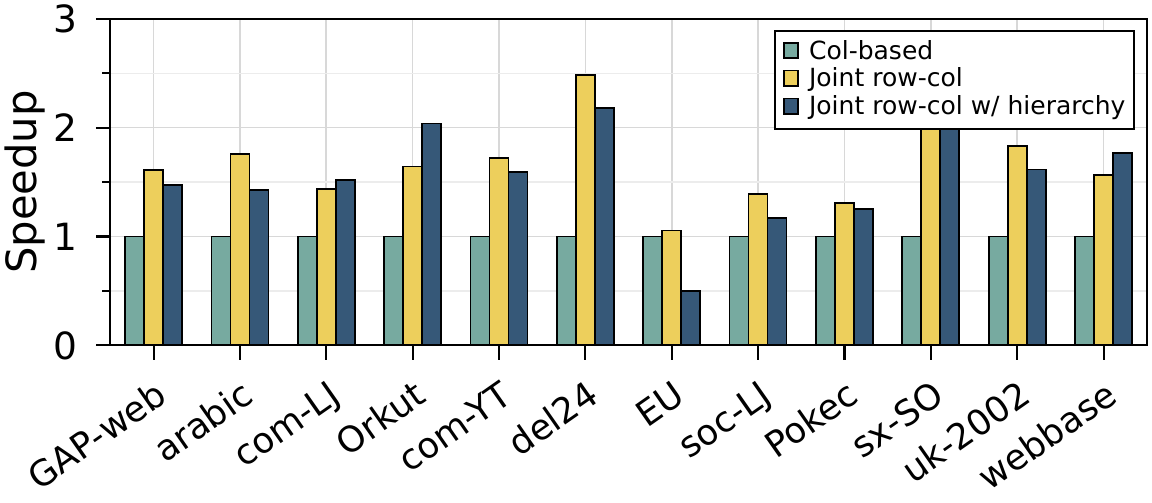}
    \vspace{-15pt}
    \caption{Step-wise optimization results on Intel GPUs, nGPUs = 24.}
    \vspace{-10pt}
    \label{fig/experiment:stepwise_intel}
\end{figure}

\subsection{Results on Intel GPUs}
\Cref{fig/experiment:stepwise_intel} shows the speedup achieved by our method on Aurora after porting the baselines to Intel GPUs. SHIRO's sparsity-aware communication remains effective on this platform, while the best-performing variant changes: Aurora's relatively balanced intra- and inter-node communication (about ~17 GB/s of inter-node bandwidth per GPU, compared with ~6 GB/s per GPU on TSUBAME4.0) makes the flat joint row-column schedule preferable to whole-node aggregation, avoiding extra packing and collective stages. This highlights that the hierarchy-aware communication strategy is most beneficial when the hardware hierarchy creates a sufficiently large bandwidth cliff; otherwise, the simpler sparsity-aware schedule can already expose the available network bandwidth. We also observed similar scaling performance, omitted for space.

\section{Related Work}

\subsection{Distributed SpMM}
{
Much prior work on distributed SpMM has focused on communication-avoiding strategies, including 1D and 2D partitionings with replication~\cite{koanantakool2016communication}. Later work focused on various algorithms for tall-skinny dense matrix multiplication~\cite{selvitopi2021distributed}, as is common for GNN and other machine learning applications, as well as asynchronous RDMA-based algorithms to avoid load imbalance issues~\cite{brock2024rdma}. 
Communication-avoiding techniques have been applied to GNNs and other domains~\cite{tripathy2020reducing,bharadwaj2022distributed}, yet primarily use sparsity-oblivious methods and exploit sparsity structures only through matrix permutation.
}

Recent work has explored sparsity-aware communication strategies for distributed SpMM. Several methods~\cite{mukhopadhyay2024sparsity, acer2016improving, zhang2025cola} proposed sparsity-aware approaches that primarily employ column-based strategies.
Abubaker et al.~\cite{abubaker2024spcomm3d} introduced a hybrid approach that combines both row-based and column-based sparsity-aware strategies. However, these methods consider row-based and column-based strategies separately, leading to communication redundancy. 
Another direction explored by~\cite{block2024two} is adaptively choosing between P2P and collective communication for different matrix blocks based on sparsity patterns, which is orthogonal to our approach. 
{Furthermore,} other work has explored using sparsity-aware techniques (column-based strategy {or coarse-grained strategy}) to optimize sparse-sparse matrix multiplication (SpGEMM)~\cite{hong2024sparse, ranawaka2024distributed}.

Another line of research reduces communication overhead through graph partitioning or matrix decomposition techniques. 
For example, prior work reduces communication through arrow matrix decomposition~\cite{gianinazzi2024arrow} or hypergraph partitioning~\cite{acer2016improving}.
These methods optimize communication patterns through matrix partitioning. Our method operates at a different level: it optimizes the communication strategy for patterns produced by any partitioning algorithm. This orthogonality allows our method to be applied on top of these partitioning schemes for additional communication overhead reduction.

Several works have focused on optimizing distributed GNN training, where SpMM-like operations constitute the primary performance bottleneck. The most related work~\cite{zhuang2025scaling} reduces communication overhead in GNN training with similar strategies, but is only tailored to graphs and does not consider hierarchical networks. Other works, such as~\cite{mukhopadhyay2024sparsity, bharadwaj2022distributed}, accelerate distributed GNN training through different approaches to reduce SpMM communication overhead. 
Some works~\cite{zhuang2025scaling, wan2022bns, wan2023adaptive} use lossy compression to reduce GNN training communication, but such methods are inapplicable to scientific SpMM that requires exact results.


\subsection{Hierarchical Communication}
{
 Hierarchical communication is a technique to reduce inter-group communication by leveraging high-bandwidth intra-group interconnects. Prior work has explored hierarchical aggregation strategies for sparse kernels in traditional HPC solvers~\cite{lockhart2023performance, bienz2018reducing, bienz2020reducing, zhang2025cola} and 3D image reconstruction systems~\cite{hidayetouglu2020petascale}. However, most of these works are typically optimized for row-based or column-based communication strategies. In contrast, our work integrates the proposed joint row-column sparsity-aware strategy with hierarchical network topology. By fully exploiting the inherent opportunities for inter-node communication reduction and complementary link utilization in the joint row-column approach, our method achieves better communication performance.
}

\section{Conclusion}
In this paper, we propose a communication-efficient distributed framework, \method{}, to address the communication bottleneck in distributed sparse matrix-matrix multiplication (SpMM). We address inefficient communication strategies by: (1) a fine-grained, sparsity-aware communication strategy that exploits sparsity patterns to eliminate redundant data transfers, and (2) a hierarchical communication strategy that leverages two-tier network architectures to minimize communication across slow network links. 
{Extensive experiments on real-world datasets demonstrate that our approach significantly reduces communication overhead, improves SpMM performance, and achieves substantial speedups in GNN training with minimal preprocessing overhead.}

\begin{acks}
We thank the anonymous reviewers for their valuable comments and suggestions. This project was supported by JSPS KAKENHI Grant Number JP25K15142 and the RIKEN Junior Research Associate Program.
\end{acks}

\bibliography{sample-base}

\scriptsize
\noindent
\newline Optimization Notice: Software and workloads used in
performance tests may have been optimized for performance only on
Intel microprocessors.  Performance tests, such as SYSmark and
MobileMark, are measured using specific computer systems,
components, software, operations and functions.  Any change to any
of those factors may cause the results to vary.  You should
consult other information and performance tests to assist you in
fully evaluating your contemplated purchases, including the
performance of that product when combined with other products.
For more information go to \url{http://www.intel.com/performance}.

\noindent Intel, Xeon, and Intel Xeon Phi are trademarks of Intel Corporation in the U.S. and/or other countries.

\normalsize

\end{document}